\documentclass[11pt]{article}
\usepackage{geometry}                
\usepackage{graphicx}
\usepackage{amssymb}
\usepackage{epstopdf}
\usepackage{graphicx,amsmath,amssymb}

\newcommand{\ba}{\begin{eqnarray}}
\newcommand{\ea}{\end{eqnarray}}
\newcommand{\be}{\begin{equation}}
\newcommand{\ee}{\end{equation}}

\title{A New View of the Cosmic Landscape}
\author{S.-H. Henry Tye \footnote{Electronic mail: tye@lepp.cornell.edu} \\
\\
Laboratory for Elementary Particle Physics \\ Cornell University, 
Ithaca, NY 14853
}

\date{}                                           

\begin{document}
\maketitle

\abstract{In this scenario, a generic meta-stable deSitter vacuum site in the cosmic landscape in string theory has a very short lifetime. Typically, the smaller is the vacuum energy of a meta-stable site, the longer is its lifetime. This view of the landscape can provide a qualitative dynamical explanation why the dark energy of our universe is so small. The argument for this scenario is based on resonance tunneling, a well-known quantum mechanical phenomenon, the topography of the landscape, and the vastness of the cosmic landscape. Mapping the topography of the landscape, even if only in a small region, will test the validity of this scenario.
}

\section{Introduction}

The discovery of dark energy reveals a crucial fact about our universe \cite{Perlmutter:1998np}. 
It leads us to believe that we are living in a vacuum state with a tiny positive cosmological constant 
$\Lambda_{0}$.
A naive cosmological constant would have a value dictated by the Newton's constant, or $10^{120}$ orders of magnitude bigger than the observed value. This is a puzzle. 

In string theory, there are 10-dimensional spacetime. To agree with observations, 6 of the spatial dimensions must be compactified into a very small size. Recent analysis of flux compactifications shows that string theory has exponentially or even infinitely many, (meta-stable) solutions \cite{Bousso:2000xa}, with a wide range of vacuum energies or cosmological constants, including ones with very small positive cosmological constants. This is referred to as the cosmic string landscape.
This is encouraging, since, if string theory is correct, our universe with a very small cosmological constant must be one of its solutions. 

In this vast landscape, any vacuum site with a positive cosmological constant is meta-stable. Although the shapes and sizes of barriers between sites are not well understood, one intuitively expects that they are of string scales and so it is argued that these sites are extremely long lived. The puzzle now becomes why we end up at a site with such a small cosmological constant, when numerous meta-stable vacua with much larger cosmological constants are present. Here I like to give a possible dynamical argument how this may happen. A key ingredient is the vastness of the landscape, precisely the property usually thought to be the origin of the puzzle. 

What properties the string landscape should have, so that a universe with a very small cosmological constant is dynamically natural ? 
Suppose the smaller is the vacuum energy of a meta-stable site, the longer is its lifetime. 
So a site with a large vacuum energy decays quickly to a small vacuum energy site with a long lifetime.
Let us go one step further. 
Suppose typical meta-stable vacuum sites in the cosmic landscape with generic cosmological constants have very short lifetimes (say, less than one Hubble time), while typical lifetimes of sites with cosmological constants  below a specific value $\Lambda_{c}$ are exponentially long (say, long compared to the 
age of our universe). In this scenario, even if the universe starts at a large cosmological constant site, 
it would decay rapidly and repeatedly if necessary until it reaches a site with a cosmological constant
below $\Lambda_{c}$. If the landscape has this property, with $\Lambda_{0} \lesssim \Lambda_{c}$, then our universe with such a small $\Lambda_{0}$ is dynamically natural. Here I like to argue that this scenario is entirely possible.

The basic observation is simple. Although tunneling from one vacuum to another typically takes an exponentially long time, it is well-known that, when the condition is right, the tunneling can be very 
efficient \cite{Merzbacher}, that is, with tunneling probability equal to unity. Consider Figure 1 in a quantum mechanical problem. 
\begin{figure}
\begin{center}
 \includegraphics[width=0.6\textwidth,angle=0]{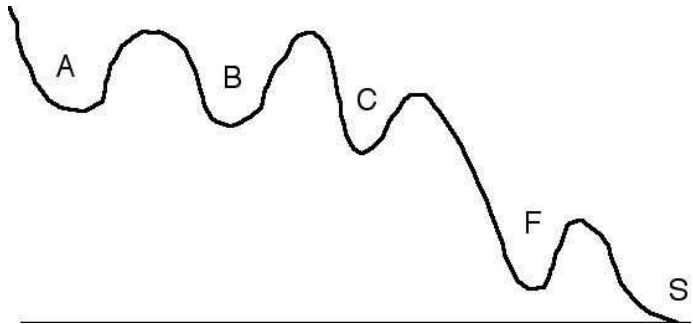} 
\caption{Tunneling from one site to another site in the cosmic landscape.  
Here, the potential $V(x)$ is plotted as a function of $x$. Although tunneling probabilities from $A$ to $B$ and from $B$ to $C$ are both exponentially suppressed, tunneling probability from $A$ to $C$ can be of order unity, a consequence of the resonance effect.}
\label{fig1}
\end{center}
\end{figure}
In general, all tunneling probabilities $T_{i \rightarrow j}$ are exponentially small. However, resonance effects can change that.
Suppose we start at site $A$. The tunneling probability $T_{A \rightarrow C} \sim 1$ if (1) the $B$ site satisfies a resonance condition, that is, the initial energy is precisely that of a bound state eigenvalue in  $B$, and (2) the exponentially small tunneling probabilities are equal : $T_{A \rightarrow B} \sim  T_{B \rightarrow C}$, which is the condition for a narrow and so sharply peaked resonance. This is the narrow resonance condition.
Together, they form the efficient tunneling condition. Note it is crucial that motion in region $B$, although classically allowed, should be treated quantum mechanically. It is the phase introduced by the particle propagation in $B$ that leads to the interference of the 2 large tunneling suppression factors. Note that efficient tunneling can also happen via a number of intermediate sites.

Another property that can be crucial is the topography of the landscape. The landscape 
around $A$ may look more like a random potential in a high-dimensional moduli space than that of a regular lattice, so presumably there are tunneling paths from $A$ to some of its neighboring sites that are barely suppressed. In a realistic scenario, the effective potential of the landscape may look regular in some directions but random in others, so this random potential feature along some directions is more likely in a higher-dimensional moduli space. 
This property of the landscape can be explicitly checked, at least for some generic meta-stable sites.

Let us now apply this property to the string landscape. Instead of a full quantum treatment of the landscape, which is beyond the scope of this paper, we emphasize the qualitative picture that emerges
when resonance tunneling effects are included.
To simplify the discussion, let us allow decays of any deSitter site in the landscape only to other sites with smaller semi-positive vacuum energies \cite{Coleman:1980aw}.  
That is, we ignore tunneling from a positive cosmological constant site to a negative cosmological constant site. (Either this does not happen, or it is followed by a big crunch, so the process is not important phenomenologically.)
In this case, a site with a large vacuum energy would have more decay channels than 
a site with a small vacuum energy.
Suppose we start at a site $A$ in the string landscape with a cosmological constant a little below the string scale. Due to the vastness of the landscape, there are exponentially many tunneling channels for $A$ to choose from. This includes tunneling from $A$ to neighboring sites as well as sites beyond via neighboring sites. Even if the probability for any tunneling channel to satisfy the efficient tunneling condition is exponentially small, one can speculate that the vastness of the landscape would enable $A$ to find at least some channels to tunnel efficiently. So $A$ decays rapidly via these channels to sites
with lower cosmological constants.
The smaller is its cosmological constant, the fewer tunneling channels it can choose from, and the less likely that the site can decay via efficient tunneling. Without resonance tunneling, a typical decay lifetime is exponentially long.
As a consequence, a site with a small cosmological constant would typically have a very long lifetime.
Sites with exponentially small cosmological constant would have so few tunneling channels open to it that in all likeliness, no efficient tunneling channels are available to them. Such sites would have exponentially long lifetimes. Presumably, we are in such a site in the string landscape. 
So the quantum landscape may look very different from the classical cosmic landscape.


If statistically, the critical vacuum energy $\Lambda_{c}$ (i.e., sites with cosmological constant below which would have no efficient channels open to them) is within a few orders of magnitude of the value of the dark energy observed, one may consider the cosmological constant problem understood. The condition for such a solution depends on the topography of the string landscape: the distribution of sites with different $\Lambda$ and the shapes and sizes of the barriers between them. In principle, these properties are calculable, certainly in the neighborhood of any specific site. That is, this scenario can be tested. If the qualitative properties of the landscape does satisfy the condition, string theory would provide a dynamical solution to the cosmological constant problem. We see that the vast string landscape plays a crucial role in this solution.

If the string landscape does have the qualitative profile required to dynamically solve the cosmological constant problem, we still need to fold these properties into the cosmological evolution of our universe. 
A proper treatment of resonance tunneling probably requires us to start with the wavefunction of the universe. It is a challenging but tractable exercise to check the validity of this scenario of the cosmic landscape.

Sec. 2 is a review of the resonance tunneling in quantum mechanics.
Sec. 3 presents the new view of the landscape. 
Sec. 4 states the necessary qualitative features of the string landscape that can provide a dynamical reason why the dark energy is so small. 
Sec. 5 contains some remarks.

\section{Narrow Resonance Tunneling}

This review follows closely that in Ref.\cite{Merzbacher}.
Consider the 1-dimensional quantum mechanical system of a particle with unit mass
\ba
S = \int dt \left[ \frac{1}{2} (\frac{dx}{dt})^{2} - V(x) \right]
\ea
where the potential $V(x)$ is shown in Figure 1.
The tunneling probability, or transmission coefficient, of a particle with energy $E$ 
is straightforward to obtain in the WKB approximation. 

Beginning with the barrier between $A$ and $B$, we are interested in the tunneling probability 
of $A$ to $B$, $\Gamma_{A \rightarrow B}$. Let the coefficients of the left- and the right-moving components of the wavefunction in $A$ be $\alpha_{L}$ and $\alpha_{R}$ respectively, and 
that in $B$ be $\beta_{L}$ and $\beta_{R}$ respectively. 
The relation between these coefficients are given by 
the WKB connection formulas,
\ba
\label{matrix1}
\left(\begin{array}{c} \alpha_{R} \\ \alpha_{L}\end{array}  
\right) 
=\frac{1}{2}\left( \begin{array}{c}
 \Theta  + \Theta^{-1}       \quad \quad \quad i\,(\Theta  -  \Theta^{-1} )\\
-i\,(\Theta - \Theta^{-1})  \quad  \quad \Theta  + \Theta^{-1}
\end{array}  
\right) 
\left(\begin{array}{c} \beta_{R} \\ \beta_{L}\end{array}  \right) 
\ea
where, in the WKB approximation,
\ba
\Theta &\simeq & 2\, \exp\left (\,\int_{x_1 }^{x_2} dx
\sqrt{2(V(x)-E)}\, \right ) ~,
\label{ThetaA}
\ea
where $x_1$ and $x_2$ are the classical turning points. 
Assuming that there is no wave incident from the right in $B$, i.e., $\beta_{L}=0$, the tunneling probability follows from the transmission coefficient from $A$ to $B$, given by
\ba
\label{tran1}
T_{A \rightarrow B} = \left| \frac{\beta_{R}}{\alpha_{R}}\right|^{2} = 
4 \left( \Theta  +\frac{1}{ \Theta} \right)^{-2} \simeq \frac{4}{\Theta^{2}} 
\ea
Generically, $\Theta$ is exponentially large, so $T_{A \rightarrow B}$ is 
exponentially small. This is the well-known WKB tunneling formula.

Next we consider the probability of tunneling from $A$ to $C$ via $B$.
The matrix relating the coefficients of the incoming wave from $A$ to $C$ is given by
\ba
\frac{1}{4}\left( \begin{array}{c}
 \Theta  +\Theta^{-1}       \quad \quad  i\,(\Theta  - \Theta^{-1}  )\\
-i\,(\Theta  -\Theta^{-1})  \quad  \Theta  +\Theta^{-1}
\end{array}  
\right)  
\left( \begin{array}{c}
e^{-i\, W} \quad 0\\
0 \quad \quad e^{i\, W} 
\end{array}  
\right)  
\left( \begin{array}{c}
 \Phi  +\Phi^{-1}       \quad \quad i\,(\Phi  -\Phi^{-1})\\
-i\,(\Phi  -\Phi^{-1})  \quad  \Phi  +\Phi^{-1}
\end{array}  
\right) \nonumber
\ea
where, in the WKB approximation, $W$ is the integral over $B$
\ba
W=\int_{x_2}^{x_3} \sqrt{2(E-V(x))}\, dx
\ea
and 
\ba
\Phi &=& 2\, \exp\left (\,\int_{x_3 }^{x_4} dx
\sqrt{2(V(x)-E)}\, \right ) 
\label{ThetaC}
\ea
where $x_3$ and $x_4$ are the respective classical turning points at the barrier between $B$ and $C$.
In general,  $T_{B \rightarrow C} = 4/\Phi^{2}$ is also exponentially small.
Now, the tunneling probability (transmission coefficient) from $A$ to $C$ via $B$ is given by
\ba
T_{A \rightarrow C} 
=4\, \left( ( \Theta \Phi + \frac{1}{\Theta \Phi})^{2} 
\cos^{2} W + \left(\Theta /\Phi+ \Phi/\Theta \right)^{2}\sin^{2} W\right)^{-1} ~.
\ea
In the absence of $B$, $W=0$ so $T_{A \rightarrow C}$ is very small,
\ba
T_{A \rightarrow C} \simeq 4 \Theta^{-2} \Phi ^{-2} =  
T_{A \rightarrow B} T_{B \rightarrow C} /4
\ea
However, if $W$ satisfies the quantum condition for bound states in $B$, 
\ba
W=(n_{B}+1/2) \pi
\label{resonancecond}
\ea
then $\cos W =0$, and the tunneling probability approaches a small but not necessarily 
exponentially small value
\ba
T_{A \rightarrow C}  = \frac{4}{\left(\Theta/\Phi+ \Phi/\Theta \right)^{2}} 
\label{resonancetnn}
\ea
This is the resonance effect. If $T_{A \rightarrow B}$ and $T_{B \rightarrow C}$ are very different, 
we see that $T_{A \rightarrow C}$ is given by the smaller of the ratios between
$T_{A \rightarrow B}$ and $T_{B \rightarrow C}$. 

Following Eq.(\ref{resonancetnn}), we see that 
\ba
T_{A \rightarrow C}   \rightarrow  1    
\ea
as
\ba
T_{A \rightarrow B} \rightarrow T_{B \rightarrow C}
\label{narrow}
\ea
We call this the narrow resonance condition. 
Together with the resonance condition (\ref{resonancecond}),
this forms the efficient resonance tunneling condition.
This means that tunneling from $A$ to $C$ passing
through an appropriate $B$ may not be suppressed at all if the efficient tunneling
condition ((\ref{resonancecond}) and (\ref{narrow})) is satisfied. 

For large $\Theta$, so that the penetration through the barriers
is strongly suppressed, the transmission coefficient has sharp
narrow resonance peaks at the values in Eq.(\ref{resonancecond}). 
Treating the resonance shape as a function of the incoming particle energy,
the resonance has a width $\Delta E$. Let the separation between neighboring 
resonances be $E_{0}$, then 
a good estimate of the probability of hitting a resonance is given by
\ba
P(A \rightarrow C) =\frac{\Delta E}{E_{0}} \simeq \frac{2}{\pi \Theta \Phi} 
\left(\frac{\Theta}{\Phi} + \frac{\Phi}{\Theta} \right)
= \frac{1}{2\pi} \left(T_{A \rightarrow B}  + T_{B \rightarrow C} \right)
\label{2steps}
\ea
We see that the probability of hitting a resonance is given by the larger of the two decay probabilities,
$T_{A \rightarrow B}$ or $T_{B \rightarrow C}$, and the average tunneling probability is given by 
\ba
<T_{A \rightarrow C}> = P(A \rightarrow C)T_{A \rightarrow C} 
\sim \frac{T_{A \rightarrow B}T_{B \rightarrow C}}{T_{A \rightarrow B} + T_{B \rightarrow C}}
\label{averageT2}
\ea
which is essentially given by the smaller of the two tunneling probabilities.

\section{The Quantum Landscape}

Now we like to apply this resonance effect  to the tunneling between different deSitter vacua in the cosmic landscape. 
In principle, we should start with the wavefunction of the universe localized at a particular site and follow its time evolution within a full quantum treatment of the landscape. (An initial wavefunction spanning the whole or a part of the landscape are also interesting possibilities.) 
However, this approach is beyond the scope of this paper. Instead, the treatment here is somewhat ad hoc. 
Let us consider a site with vacuum energy below the string scale so a supergravity approximation is valid around this site. 

(1) For a single modulus $\phi$, let us simplify the problem by considering only its time-dependence,
$x(t)=\phi(t)$. In this case, $V(\phi) \rightarrow V(x)$ and the above analysis applies. 
More generally, the string landscape involves a set of moduli and other light scalar modes $\phi_{i}$, with an approximate Langrangian density 
\ba
L_{Einstein} -\frac {1}{2} G^{ij}(\phi) \partial_{\mu}\phi_{i} \partial^{\mu} \phi_{j}  - U(\phi_{i})
\ea
Again, let us consider only the time-dependence of $\phi_{i}$, that is, $\phi_{i} \rightarrow \phi_{i}(t)$. 
For $G_{ij} = \delta_{ij}$, this becomes a multi-dimensional quantum mechanical problem. The tunneling from $A$ to $C$ via $B$ involves a set of paths. More generally, let $x$ is the path length 
\ba
x = \int^{t} dt' \sqrt{G^{ij}(d\phi_{i}/dt) (d\phi_{i}/dt)}
\ea
so that $G^{ij}(d\phi_{i}/dt) (d\phi_{i}/dt)  \rightarrow (dx/dt)^{2}$. For a fixed path length $x$, there are many paths in the field space. For each path, $U(\phi_{i}) \rightarrow V(x)$; that is, a different path yields a different $V(x)$. In terms of $x$ and $V(x)$, the above resonance tunneling properties in quantum mechanics can be applied here. In practice, we expect the physics to be dominated by a specific path, which may be found via the variational method. 

(2) Note that the bounce treatment in Ref.\cite{Coleman:1980aw} treats $x$ classically in the classically allowed region ($B$ here), while the resonance effect shows up only if $x$ is treated quantum mechanically even when it is in the classically allowed region. This difference is crucial here. 

(3) To simplify the problem, we shall ignore tunneling to sites in the landscape with negative cosmological constants.
Either tunneling from a positive cosmological constant site to a negative cosmological constant site
is forbidden, or that resulting site ends in a big crunch so it plays no important role here \cite{Coleman:1980aw}. If the tunneling followed by a big crunch happens, this simply enhances the decay width
the meta-stable dS sites.

(4) To simplify the discussion further, we shall ignore tunneling from any site $A$ to sites with vacuum energies larger than $\Lambda_{A}$. The tunneling probability of any given meta-stable (deSitter) site $A$ should be dominated by tunneling to sites $C$ where the vacuum energy $\Lambda_{C} \le \Lambda_{A}$. Tunneling up is so unlikely that it should be safe to ignore them. This is consistent with our usual understanding.

(5) Depending on the details such as cosmological evolutions, one may like to tighten or relax the efficient tunneling condition. To simplify the discussions, cosmological issues such as inflation and big bang are not taken into account here. In general, efficient resonance tunneling requires that $T_{A \rightarrow B}$ and $T_{B \rightarrow C}$ to be within comparable orders of magnitude. 

(6) In the quantum mechanical problem discussed in Sec. 2, the average tunneling probability
is for a particle tunneling from $A$ to $C$ with different energies (or a broad wave packet). We may re-interpret the average tunneling probability (\ref{averageT2}) as that with fixed energy but tuning the shape of $B$. 
In the application to the landscape here, we are considering the tunneling of the one and only wavefunction of the universe from $A$ to different sites $C$s via different $B$s. Here, the average is over all tunneling channels from $A$ to different $C$s. So the properties in the quantum mechanical problem is only qualitatively meaningful here. Note that $B$ is a classically allowed region. It can be a ``valley'' between $A$ and $C$.

(7) Resonance tunneling can happen repeatedly as the universe moves towards lower vacuum energy sites. It is more appropriate to treat the process as a single tunneling via a series of sites. 
That is, we should consider tunneling channels  like 
$A \rightarrow B_{1} \rightarrow B_{2} \rightarrow . .  . . . \rightarrow B_{n-1} \rightarrow C$, where the evolution in $C$ may be treated classically. The tunneling probability of such a channel is
$T_{A \rightarrow C}(n)$, where $n \ge 2$ is the number of steps involved. 
In this case, the tunneling probability depends on all the sites $B_{i}$ and the barriers between them
and $A$ and $C$. Without the resonance effect, we expect the naive tunneling probability to be
\ba
{\hat T}_{A \rightarrow C}(n) \simeq T_{A \rightarrow B_{1}} T_{B_{1} \rightarrow B_{2}} . . . .
 T_{B_{n-1} \rightarrow C}
\label{naiveT}
\ea
that is, the product of the individual tunneling probabilities. 
When all $T_{i \rightarrow j}$ are comparable, i.e., 
$T \simeq T_{i \rightarrow j}$, then ${\hat T}_{A \rightarrow C}(n) \sim T^{n}$.
For comparison, it is useful to get an order of magnitude estimate of the probability of efficient 
tunneling from $A$ to $C$ when the resonance effect is taken into account.

The resonance condition (\ref{resonancecond}) is simply an interference that leads to the cancellation of the large suppressions in the tunneling probability. This same effect can take place in a multi-step tunneling channel. Let the phase in region $B_{i}$ be $W_{i}$. 
We note that the resonance condition is co-dimenion one
in the $(n-1)$-dimensional $W_{i}$-space. For the $n=2$ case discussed in Sec. 2, the resonance conditions (\ref{resonancecond}) are  points along the $W$ line. For $n=3$, the resonance conditions
are curves in the $W_{1}-W_{2}$ space, and so on. That is, satisfying the resonance condition requires
only one condition among the $W$s. This means that the probability of hitting a resonance is not suppressed in the $n$-step tunneling. However, the probability of reaching efficient tunneling decreases. 

Crudely speaking,
\ba
<T_{A \rightarrow C}(n)> = P(A \rightarrow C) T_{A \rightarrow C} \sim  
{Min} \left(T_{min},  \sqrt{{\hat T}_{A \rightarrow C}(n)} \right)
\label{realrate}
\ea
where $T_{min}$ is the smallest of the individual $T_{i \rightarrow j}$ and $<T_{A \rightarrow C}(n)>$ is roughly equal to $T_{min}$ or the square root of the products of the individual tunneling probabilities,
whichever is smaller.
Although this is still exponentially small, it is exponentially bigger than that in the case where resonance effects are ignored, $\hat T_{A \rightarrow C} \sim T^{n}$ (\ref{naiveT}).
So the probability to have an efficient tunneling via $n$ steps, in the case where all individual $T_{i \rightarrow j}$ are of comparable orders of magnitude, is crudely given by, for each tunneling path,
\ba
P(A \rightarrow C) \sim \left( T_{A \rightarrow B_{1}} . . . .T_{B_{n-1} \rightarrow C} \right)^{1/2}  \quad  \quad n \ge 2
\label{nsteps}
\ea
Here, $n=2$ reproduces Eq.(\ref{2steps}). 

Now we are ready to discuss the tunneling probability of any site $A$ in the string landscape. The tunneling probability of any given meta-stable site $A$ is simply the sum of its tunneling probability
to any other site $C$, that is $T_{A} = \sum_{C} T_{A\rightarrow C}$, where the sum is over sites 
with $0 \le \Lambda_{C} < \Lambda_{A}$. 
Since typical tunneling probabilities are exponentially suppressed, so $T_{A} \sim 1$ only if there is at least one tunneling channel satisfying the efficient tunneling condition. The number of efficient tunneling channels available to $A$ may be estimated to be
\ba
N(A) = \sum_{C} P(A \rightarrow C)  \quad  \quad \Lambda_{C} < \Lambda_{A}
\label{numbertunnel}
\ea
where $P(A \rightarrow C)$ is the probability that the tunneling $A \rightarrow C$ is efficient 
(for any $n \ge 2$). If the string landscape has an infinite number of meta-stable sites, then there is no guarantee that this sum converges. However, the estimate for $P(A \rightarrow C)$ (\ref{nsteps}) above  suggests that contributions from sites far from site $A$ are heavily suppressed so the sum is likely to converge. We shall assume this is the case for all sites we are interested in in the landscape.

Although each $P(A \rightarrow C)$ is typically exponentially small, the sum over $C$ covers the whole landscape, so the resulting $N(A)$ needs not be small. The number of channels with just $n=2$ steps is probably exponentially large due to the large number of moduli and other light scalar modes, which can number from dozens to hundreds. (We expect the wavefunction to be localized if the number of moduli is small.) For a fixed tunneling channel, there are many paths $x$ can take. This may provide an enhancement to $T_{A \rightarrow C}$ and so $P(A \rightarrow C)$. It would be important to estimate this. Furthermore, the number of tunneling channels available should grow rapidly as $n$ increases. 

If $N(A) > 1$, then $T_{A} \sim 1$ and this classically meta-stable site is actually very unstable.
In fact, its decay may be so fast that eternal inflation is absent. This would be the case if its lifetime is shorter than the corresponding Hubble time. 
If $N(A) < 1$, then no efficient tunneling is likely and $T_{A}$ would be exponentially small. Such a site would have exponentially long lifetime.

Since the upper bound of the sum over all sites is cut off by $\Lambda_{A}$, we expect $N(A)$ to be sensitive to $\Lambda_{A}$. Because the number of tunneling channels contributing to the sum in $N(A)$ decreases as $\Lambda_{A}$ decreases, one may conjecture that, statistically at least, sites with large $\Lambda$s tend to decay much more rapidly compared to sites with small $\Lambda$s. This view of the landscape is quite different from the usual picture.  

\section{Conditions for a Solution to the Cosmological Constant Problem}

In principle, more detailed studies of the string landscape will yield the topography of the landscape around any site : the distribution of sites and their vacuum energies as a function of the path length $x$
as well as the barriers between them. One can then check the convergence of the sum in Eq.(\ref{numbertunnel}) and evaluate $N(A)$. Using these informations, we can estimate how likely our universe would end up in some specific sites. 
This is a challenging but not insurmountable problem.
Since this information is not yet available, let us discuss the properties of the landscape that would lead to our today's universe, that is, a meta-stable universe with an exponentially small cosmological constant and exponentially long lifetime. 

The number of channels contributing to the sum in $N(A)$ depends on the value of $\Lambda_{A}$, where the number of channels approaches zero as $\Lambda_{A} \rightarrow 0$. 
(As an illustration, $N(A) \sim N_{0}\Lambda_{A}^{k}$ with $k>0$ and a large enough $N_{0}$ would do.)
A site with exponentially many available tunneling channels would decay rapidly while a site with limited  number of tunneling channels would be very long lived.
Statistically, there is a critical value of $\Lambda_{c}$ such that $N(A)=1$ for 
$\Lambda_{A}=\Lambda_{c}$. That is, $N(A) >1$ for $\Lambda_{A} > \Lambda_{c}$, and
 $N(A) <1$ for $\Lambda_{A} < \Lambda_{c}$. 
Sites with $\Lambda > \Lambda_{c}$ would decay relatively efficiently while sites with $\Lambda < \Lambda_{c}$ would be long lived. 
That is, there is a sharp jump in the lifetimes as $\Lambda$ decreases past $\Lambda_{c}$. If $\Lambda_{c}$ turns out to be close to the value of the dark energy in our universe, one may consider this as a qualitative explanation of the cosmological constant problem.

Let us now consider the wavefunction of the universe. One may start by considering a wavefunction that is localized to a specific classically allowed site, say $A$, with a generic $\Lambda$. The decay width of $A$ is given by
\be
\Gamma_{A} \simeq M \sum_{C}  <T_{A \rightarrow C}(n)>
\ee
where $ <T_{A \rightarrow C}(n)>$ is given in Eq.(\ref{realrate}) and $M$ is the mass scale relevant here. Some reasonable choices are $M \sim m_{s}$, $M \sim H \sim \Lambda_{A}^{1/2}/M_{Pl}$ or 
$M \sim \Lambda_{A}^{1/4}$. With either of the latter choices, we see that the lifetime of $A$ increases as a function of $\Lambda_{A}$.

The landscape in the neighborhood of a generic site $A$ is probably closer to a random potential in a $k$-dimensional moduli space than a regular lattice. In this situation, some of the tunneling paths from $A$ are not suppressed at all. As in Anderson localization, one may expect that the wavefunction rapidly spreads beyond $A$, but is still localized in some neighborhood around $A$. This localization of the wavefunction may still allow it to spread over a number of classically stable sites with $\Lambda$ comparable to or smaller than that of $A$. Suppose the correlation length of the wavefunction is larger than the typical separation between such sites, say by a factor of $r>1$. Then the wavefunction is spread over $r^{k}$ sites. If any of these sites in the patch around $A$, say $D$, has a substantially smaller $\Lambda$ than other neighboring sites of $A$, the wavefunction will tend to become localized there (via decoherence or otherwise). Next, the wavefunction will spread around the neighborhood of this vacuum site $D$, where neighboring sites with $\Lambda$ smaller than or comparable to that of $D$ will contribute. This process may repeat quickly so that the wavefunction may end up at a small $\Lambda$ site, say $F$. This rapid process will stop at site $F$ if the neighborhood of $F$ has no site (or very few sites) with comparable or  smaller $\Lambda$ so that tunneling away from $F$ is exponentially suppressed; that is, $r \le 1$.


If the landscape does satisfy this property, then we have the following scenario.  Suppose we start
at a site with $\Lambda$ close to but below the string scale. It will decay rapidly to another site with a lower $\Lambda$. This process may repeat any number of times. (Here, the tunneling may still allow a site to inflate some number of e-folds before decay.) Finally we reach a site $F$ (in Figure 1) where $N(F) <1$ and $\Lambda_{F} < \Lambda_{c}$. Because of the limited number of sites it can decay to, the decay time of $F$ to another lower $\Lambda$ site, say a supersymmetric site $S$ with zero $\Lambda$, would be exponentially long. When this happens, we expect $F$ to have a very long lifetime. Presumably, this is the site our universe is living in today. That is, it does not matter at which site the universe starts; in a short time, our universe would have ended in a site with a very small cosmological constant and a long lifetime.

\section{Remarks}

In the nucleation bubble picture in the thin wall approximation \cite{Coleman:1980aw}, 
bubbles of all sizes keep popping out, a consequence of quantum fluctuations. However, because the domain wall surface tension overcomes the volume effect, classically, small bubbles shrink to zero and play no role in the tunneling. Only large enough bubbles would be able to grow classically and complete the tunneling process. Since  large bubbles are much less likely to be created, the typical resulting tunneling rate is exponentially suppressed. Although it is not clear how resonance tunneling happens within this picture, presumably tiny bubbles survive and propagate quantum mechanically in the classically allowed region and contribute to the resonance tunneling process. These tiny bubbles would collide and release the vacuum energy difference into radiation and/or matter.

Let us consider the vacuum site $B$.
For large values of $\Lambda$ below the Planck mass $M_{P}$, quantum effect in the classically allowed region can be important. A deSitter vacuum has a finite number of degrees of freedom and a horizon. A typical energy spectrum has level (band structure) spacing of order the Hubble scale 
$H = \sqrt{\Lambda}/M_{P}$ in 4-dimensional spacetime. Hitting an energy level would allow resonance tunneling from $A$ to $C$ via $B$. As $\Lambda$ decreases, the energy level spacing decreases. When the energy spacing is small and the spectrum is dense, quantum effects become unimportant. This would shut off the resonance tunneling via $B$. This may be happening when we consider the tunneling from $F$ with a very small cosmological constant, since $\Lambda_{B} \le \Lambda_{F}$. This may provide another reason why the lifetime of a site like our today's universe is very long.

In a cosmological context, the tunneling from one site to another site with a lower cosmological constant is generically accompanied by some radiation. Suppose our universe has just tunneled to site $A$ with some radiation. If resonance tunneling is enabled when the radiation has a specific density, resonance tunneling would happen as the universe expands and the radiation is red-shifted towards zero, passing through the specific densities required for efficient tunneling. 

With the tunneling probability $T \sim 1$, the decay width $\Gamma$ is expected to be somewhere between the (warped) string scale $m_{s}$ and $H$, where we expect $m_{s} > H$. 
For $\Gamma \sim m_{s}$, there is no eternal inflation when the site can decay rapidly.
For $\Gamma \sim H$, eternal inflation may be absent as well. (In the presence of sufficient radiation, the universe expands but not inflates.) Of course, a decrease of $T$ would allow eternal inflation. For $\Gamma$ comparable to $H$, the viability of the old inflationary scenario should also be re-examined.

If the string landscape does have the qualitative profile required to dynamically solve the cosmological constant problem, we can appreciate string theory in this new light : it provides a vast landscape so that a small cosmological constant vacuum is among its numerous semi-classical solutions, and the same vastness of the landscape destabilizes all vacua except ones with an exponentially small cosmological constant, thus allowing a universe like ours to emerge, survive and grow.

\vspace{5mm}

I thank Philip Argyres, Tom Banks, Xingang Chen, Hassan Firouzjahi and Sash Sarangi for discussions.
This work is supported by the National Science Foundation under grant PHY-0355005.

\vspace{5mm}



\begin{thebibliography}{}

\bibitem{Perlmutter:1998np}
  B.~P.~Schmidt {\it et al.}  
  Astrophys.\ J.\  {\bf 507}, 46 (1998)
  [arXiv:astro-ph/9805200]; \\
  ~A.~G.~Riess {\it et al.} 
  Astron.\ J.\  {\bf 116}, 1009 (1998)
  [arXiv:astro-ph/9805201]; \\
  ~ S.~Perlmutter {\it et al.}  
  Astrophys.\ J.\  {\bf 517}, 565 (1999)
  [arXiv:astro-ph/9812133].
  
\bibitem{Bousso:2000xa}
  R.~Bousso and J.~Polchinski,
  JHEP {\bf 0006}, 006 (2000)
  [arXiv:hep-th/0004134]; \\
  S.~Kachru, R.~Kallosh, A.~Linde and S.~P.~Trivedi,
  Phys.\ Rev.\ D {\bf 68}, 046005 (2003)
  [arXiv:hep-th/0301240].

\bibitem{Merzbacher}
See e.g., E.~Merzbacher, Chapter 7 in {\it Quantum Mechanics}, 2{\it nd} edition, John Wiley, 1970.

\bibitem{Coleman:1980aw}
  S.~R.~Coleman and F.~De Luccia,
  Phys.\ Rev.\ D {\bf 21}, 3305 (1980).
  
  
  \end{thebibliography}
\end{document}